\documentclass[twocolumn,showpacs,preprintnumbers,amsmath,amssymb]{revtex4}

\usepackage{graphicx}
\usepackage{dcolumn}
\usepackage{bm}

\begin{document}

\preprint{APS/123-QED}

\title{Manuscript Title:\\with Forced Linebreak}

\title{Non-Gaussian Signatures in the Lens Deformations of the CMB Sky.
A New Ray-Tracing Procedure}

\author{P. Cerd\'a-Dur\'an, V. Quilis, and D. S\'aez}

\email{diego.saez@uv.es}
\affiliation{%
Departamento de Astronom\'{\i}a y Astrof\'{\i}sica, Universidad de Valencia,
46100 Burjassot, Valencia, Spain\\
}%


\date{\today}

\begin{abstract}
We work in the framework of an inflationary 
cold dark matter universe with cosmological constant, in 
which the cosmological inhomogeneities are
considered as gravitational
lenses for the CMB photons.
This lensing deforms the angular distribution of the CMB maps 
in such a way that 
the induced deformations
are not Gaussian.
Our main goal is the estimation of
the deviations with respect to Gaussianity appeared 
in the distribution of deformations.
In the new approach used in this paper, 
matter is evolved with a particle-mesh N-body code and,
then,
an useful ray-tracing technique designed to calculate 
the correlations of the lens deformations 
induced by nonlinear structures is applied.
Our approach is described in detail and tested.
Various correlations are estimated at an
appropriate angular scale. The resulting values point out
both deviations with respect to Gaussian statistics and
a low level of correlation in the lens deformations.
\end{abstract}

\maketitle

\section{\label{sec1}Introduction}

In the absence of a reionization, 
the photons of the Cosmic Microwave Background (CMB)
are not scattered
from decoupling to present time,
except inside galaxy clusters
(Sunyaev-Zel'dovich effect); nevertheless,
the propagation
direction of these photons changes due to 
the gravitational action
of the cosmological inhomogeneities (lensing). 

The so-called primary CMB anisotropies were produced by 
linear structures at high redshifts. 
The corresponding temperature contrast,  
$\Delta_{_{P}} \equiv  (\Delta T/T)_{_{P}}$, is a statistical field which  
can be expanded in spherical harmonics as follows:
$\Delta_{_{P}} = \sum_{m = -\ell}^{+\ell} a_{\ell m} Y_{\ell m}$,
where 
the $a_{\ell m}$ quantities are 
statistically independent variables
depending on the sky realization we are expanding.
If many sky realizations are averaged,
the resulting $a_{\ell m}$ coefficients have zero means and variances 
$\langle \mid a_{\ell m} \mid^{2} \rangle
= C_{\ell}$.
Since $\Delta_{_{P}}$ 
is a homogeneous and isotropic statistical field,
quantities $\langle \mid a_{\ell m} \mid^{2} \rangle$ 
do not depend on $m$.
In the model under consideration, scalar
energy density fluctuations are initially Gaussian, and
they remain 
Gaussian during linear evolution, 
by this reason, the primary
CMB anisotropies --which were produced by linear inhomogeneities-- 
also are Gaussian.

Gaussian primary anisotropy appears to be superimposed to
secondary anisotropies generated well
after decoupling, and also to the anisotropy of the 
microwave radiation
emitted by both our galaxy and some extragalactic sources
(contaminant foregrounds). 
These foregrounds are 
sub-dominant non Gaussian components contributing to the 
total observable signal at microwave wavelengths.
In the model we are considering here, 
there are various secondary anisotropies. A few comments about 
them are worthwhile: 
the late integrated Sachs-Wolfe anisotropy
was produced by quasi-linear inhomogeneities \cite{ful00}
and, consequently,
deviations with respect to Gaussianity are not expected to
be very important; another 
secondary gravitational anisotropy was
produced by strongly nonlinear structures as galaxy clusters and
substructures (Rees-Sciama effect), this component is sub-dominant 
and non Gaussian, and its deviations with respect to Gaussianity 
were studied by various authors
\cite{mol95,mun95,ali02}; 
finally, the Sunyaev-Zeldovich anisotropy
was produced by hot plasma in galaxy clusters, and it 
is also non Gaussian \cite{coo01}.

The main goal of this paper is the analysis of 
the non Gaussianity caused by lensing, and such an analysis
can be performed using the operative ray-tracing procedure
designed in next sections.
In the absence of lensing, there are directional 
frequency shifts which cause an anisotropic 
temperature distribution in the CMB.
The lens effect
does no cause frequency shifts, but it produces
angular deviations of the propagation directions.
This means that the lens effect does not produce any
alteration of the CMB temperature. It changes
the propagation direction 
of the CMB photons and, then,
the angular distribution of the CMB temperature 
changes accordingly. 
In other words, the lens effect deforms the
maps of the CMB temperature which appear
as a result of pure frequency shifts.
Any numerical estimate of the alterations of the CMB maps produced 
by lensing (including non Gausianity generation)
involves three steps: (i) maps of the CMB temperature distribution 
are built up for the model under consideration (without any 
lensing), (ii) these maps are deformed
taking into account the deviations of the
propagation directions produced by lensing, and (iii)
the resulting deformations of unlensed maps (hereafter called either
lens deformations or lens effect) 
are analyzed.
Let us now consider each of these steps in more detail.

In order to built up unlensed maps of the CMB, we assume
that, in the 
absence of lensing, the Gaussian primary anisotropy
dominates and, consequently, the maps to be deformed can be 
built up using
the angular power spectrum of this primary anisotropy, 
which has been calculated using CMBFAST \cite{seza96} for the 
$\Lambda$CDM model under consideration.
As it is discussed below, only small Gaussian maps having 
sizes of a few degrees and resolutions of the order of one
arc-minute are necessary and they are built up
using the Fourier Transform (FT) and the mentioned spectrum
($C_{\ell}$ quantities).
See \cite{sae96} and references cited therein
for details.
The resulting maps have spots with different sizes 
and amplitudes (angular structure). Spots with 
angular size $\pi / \ell $ have a mean amplitude 
proportional to $C_{\ell}^{1/2}$; hence, spots with 
angular size close to one degree ($\ell \sim 200$)
have large amplitudes, whereas spots corresponding to
one arcminute ($\ell \sim 10000$) have negligible amplitudes
as a result of the smallness of $C_{10000}$. For the 
spectrum under consideration,
significant spots have sizes of various arcminutes.

Cosmological inhomogeneities produce lensing and this effect is
associated to the deviation field 
$\vec {\delta}$. 
The anisotropy observed in the 
$\vec {n}$ direction is the primary anisotropy 
corresponding to the direction 
$\vec {n}_{0}$ in the absence of lensing, where
$\vec{n} = \vec {n}_{0} - \vec {\delta} $;
hence,
\begin{equation}
\Delta (\vec {n}) = \Delta_{_{P}} (\vec {n}_{0}) =
\Delta_{_{P}} ( \vec {n} + \vec {\delta} )  \ .
\label{defor}
\end{equation}
The unit vectors 
$\vec {n}$ and $\vec {n}_{0}$ point towards two points
of the last scattering surface, and the deviation field 
$\vec {\delta}$ gives the angular excursion on this surface
due to lensing.
We must estimate the deviation field 
$\vec {\delta} $ (see below) to get the deformed 
observable map $\Delta$.
If the $\vec {\delta} $ deviations are smaller than the 
size of the smallest significant spots (various minutes),
Eq. (\ref{defor}) can be expanded in $\vec {\delta} $
powers to get 
\begin{equation}
\Delta (\vec{n}) \simeq 
\Delta_{_{P}} (\vec{n})
+ \frac {\partial \Delta_{_{P}}}{\partial \vec{n}} 
\cdot \vec{\delta} 
\label{dmap}
\end{equation}
We see that map deformations, namely, the differences in
the temperature contrasts corresponding to associate directions
(second term of the r.h.s.)
are the product of two statistical fields:             
$\Delta_{_{P}}$ and $\vec {\delta}$ and, even if these field are 
Gaussian and statistically independent, the resulting product
would be non Gaussian.  

Several authors have studied angular correlations in lensed
CMB maps. In the most theoretical papers on this subject, 
some correlations have been estimated without using simulations; 
the approach used in these papers only applies for those correlations 
which can be written in terms of the time varying form of the 
matter power spectrum. In the most basic of these papers 
\cite{sel96}, the mentioned spectrum was modelled beyond 
the linear regime and the second order correlations of
lensed CMB maps were estimated.
A similar method was used by Bernardeau \cite{ber97} to estimate fourth order
correlations corresponding to various sets of four
directions (with distinct configurations); furthermore, 
this last author pointed out that third order 
correlations vanish. Other methods are necessary
to estimate correlations of order $n$ for $n>4 $, and also
to simulate CMB maps with the true 
statistics induced by lensing, namely, maps having the true 
correlations at any order (not only at second and fourth orders).
The most powerful technique to study lens distortions in CMB maps
is the use of {\em ray-tracing through N-body simulations}.
Two methods to apply this tecnique
are described in references \cite{jai00} and \cite{whi01}
(other relevant 
papers in this field are cited in these two references). 
Another new method is proposed here. 
We agree with White \& Hu \cite{whi01}, who stated that: 
{\em ..... On subdegree scales, a full description of 
week lensing therefore requires numerical simulations, the most natural 
being N-body simulations ......};  
in fact, using ray-tracing through 
N-body simulations, CMB photons are deviated by fully nonlinear 
structures (as galaxy clusters) which move in the simulation box.
Nowadays, only N-body simulations lead to a proper description of
a distribution of strongly nonlinear structures.
Unfortunately, these simulations constraint us to work in a 
periodic universe, and 
relevant problems associated to periodicity must be solved. 
Each method --designed to use ray-tracing through N-body simulations--
corresponds to a different way
for preventing periodicity effects. In this paper, one of these methods
is described and tested.

Hereafter, quantity  h is the reduced Hubble constant
$h=10^{-2}H_{0}$, where $H_{0}$ is the 
Hubble constant in units of $Km s^{-1} Mpc^{-1}$, the
density parameters corresponding to baryons,  dark matter, and  
vacuum, are  $\Omega_{b}$, $\Omega_{d}$ and  $\Omega_{\Lambda}$,        
respectively,
the total density parameter is
$\Omega = \Omega_{b} + \Omega_{d} + \Omega_{\Lambda}$, and
the matter density parameter is 
$\Omega_{m} = \Omega_{b} + \Omega_{d}$.
In the flat inflationary universe under consideration,
the above parameters take on the following values:
$h=0.65$,
$\Omega_{b}=0.05$, $\Omega_{d}=0.25$, and $\Omega_{\Lambda}=0.7$
and, then, 
according to Eke et al. \cite{eke96}, 
the power spectrum of scalar energy density
perturbations must be normalized with $\sigma_8=0.93$ in 
order to have cluster abundances compatible with observations.
Units are chosen in such a way that 
$c=8\pi G =1$, where $c$ is the
speed of light and $G$ the gravitation constant.
Whatever quantity "$A$" may be, $A_{_{L}}$ and $A_{0}$ stand for
the $A$ values on 
the last scattering surface and at present time,
respectively. The scale factor is $a(t)$, where $t$ is the
cosmological time, and its present
value, $a_{0}$, is assumed to be unity, which is always possible 
in flat universes.

\section{\label{sec2}Formalism}

The CMB photons move on null geodesics
and the line element is:
\begin{equation}
ds^{2}=-(1+2 \phi)dt^{2}+(1-2 \phi)a^{2}\delta_{ij}dx^{i}dx^{j}
\label{lel}
\end{equation}
where function $\phi$ satisfies the 
equation:
\begin{equation}
\Delta \phi = \frac {1}{2} a^{2} (\rho - \rho_{m0}) \ ,
\label{cden}
\end{equation}
and $\rho_{m0} = \Omega_{m0} \rho_{crit}$ 
is the background energy density for matter.
On account of this last equation, function $\phi$ can be 
interpreted as the peculiar gravitational potential created 
by the cosmological structures.
Equations (\ref{lel}) and (\ref{cden}) are valid for linear inhomogeneities 
located well inside the horizon and also for nonlinear
structures (potential approximation), see 
\cite{mar90}.

The deviation field $\vec{\delta}$ is given by 
the following integral \cite{sel96}:
\begin{equation}
\vec{\delta} = -2 \int_{\lambda_{_{L}}}^{\lambda_{0}} 
W(\lambda) \vec {\nabla}_{\bot } \phi \ d \lambda \ ,
\label{devi}
\end{equation}
where $\vec {\nabla}_{\bot } \phi = - \vec{n} \wedge \vec{n}
\wedge \vec{\nabla} \phi$
is the transverse gradient of the potential, and
$W(\lambda) = (\lambda_{_{L}} - \lambda)/ \lambda_{_{L}} $.
The variable $\lambda$ is 
\begin{equation} 
\lambda (a) = H_{0}^{-1} \int_{a}^{1} \frac {db} {(\Omega_{m0}b+
\Omega_{\Lambda} b^{4})^{1/2}} \ .
\end{equation}                
The integral 
in the r.h.s. of Eq. (\ref{devi})
is to be evaluated along the background null geodesics.
In our flat background,
the equations of the null geodesics passing by
point $x^{i}_{_{P}}$ are: 
\begin{equation} 
x^{i}=x^{i}_{_{P}} + \lambda (a) n^{i} \ ,
\label{ng}
\end{equation}

In next section, 
spherical clusters
with Navarro-Frenk-White (NFW) density profiles \cite{nav96}
are considered for qualitative analysis. 
These profiles appear in N-body simulations 
of dark matter halos and they have the form:
\begin{equation}
\rho(r) = \frac {\delta_{c} \ \rho_{m0} } 
{(r/r_{s})[1+(r/r_{s})]^{2}}   \ ,
\label{NFW}
\end{equation}
where $\rho_{m0} = \rho_{crit} \Omega_{m0}$,  and
$\rho_{crit} = 3H^{2}$ is the critical density of the
universe. The parameters 
$\delta_{c}$ and $r_{s}$ are related to both
the concentration parameter $c_{_{NFW}}$
and the mass $M_{200}$. This mass is that contained
inside a sphere whose mean density is $200$.
The radius of this sphere is denoted 
$r_{200}$. The following relations hold:
\begin{equation}
r_{s}=(3M_{200}/800 \pi \rho_{m0})^{1/3}/c_{_{NFW}} 
\end{equation}
\begin{equation}
\delta_{c} = \frac {200c_{_{NFW}}^{3}} {ln(1+c_{_{NFW}})
-c_{_{NFW}}/(1+c_{_{NFW}})}     \ .
\end{equation}
The profile (\ref{NFW}) remains almost unchanged from
virialization ($z \sim 1$) to present time. Some authors
have tried to assume the NFW profile with variable 
values of the involved parameters to go beyond
virialization --until $z\sim 5$ in reference \cite{bul01} --.
Other authors use generalized profiles \cite{wyi01}. 
Fortunately, our qualitative 
analysis --presented in next section-- does not require
any detailed z-dependent profile.

In Sec. III, three clusters
--which are hereafter called  $C_{_{I}}$, 
$C_{_{II}}$, and $C_{_{III}}$-- are considered.
Cluster $C_{_{I}}$ is a rich one having
$M_{200} = 1000 \ M_{g}$ 
(where 
$M_{g}=10^{12} \ M_{\odot }$) 
and $c_{_{NFW}} = 6.8$,
cluster $C_{_{II}}$ is a standard cluster having a few cents
of galaxies with $M_{200} = 250 \ M_{g}$ and $c_{_{NFW}} = 7.7$,
and cluster $C_{_{III}}$ is a galaxy group having a few tens of
galaxies with  
$M_{200} = 25 \ M_{g}$ and $c_{_{NFW}} = 12.5$. 
Since the NFW profile diverges as
physical radius $r$ tends to zero, 
an uniform core of radius  $r_{c}$ has been assumed and,
consequently, the density profile has the form 
(\ref{NFW}) for $r \geq r_{c}$ and a constant value
for $r \leq r_{c}$. Both profiles match continuously at 
$r = r_{c}$.
For this mass distribution, function $d \phi /dr$ has the 
following form:
\begin{equation}
\frac {d \phi } {dr} =\frac {1}{6} \rho(r_{c}) r
\label{dp1}
\end{equation}
for $r \geq r_{c}$, where $\rho(r_{c})$ is given by Eq. 
(\ref{NFW}), and
\begin{equation}
\frac {d \phi } {dr} = \frac {1} {r^{2}} [ A+Dr_{s}^{3} 
\{E(r)-E(r_{c})\}]  
\label{dp2}
\end{equation}
for $r \leq r_{c}$,
where $A=(1/6)r_{c}^{3}\rho(r_{c})$, $D= \delta_{c} \ \rho_{m0} /2$,
and $E(r) = ln [1+(r/r_{s})] + [1+(r/r_{s})]^{-1} $. Core
radius of $0.2 \ Mpc$, $0.1 \ Mpc$, 
and $0.05 \ Mpc$ are assigned to clusters 
$C_{_{I}}$, $C_{_{II}}$, and $C_{_{III}}$, respectively.
In next section, our conclusions are probed to be 
highly independent on this assignation.

Spherical linear inhomogeneities evolving inside the
effective horizon are also considered in
Sec. III for qualitative discussion.
These spherical regions are assumed to be uniform 
and they are characterized by their
comoving radius $R_{_{C}}$. This radius fixes the associated 
comoving scale $k_{_{C}}= \pi / 2R_{_{C}}$ and, then, the 
present density contrast is fixed using the power spectrum of the
energy density perturbations --in the model under
consideration-- which gives the average amplitude of the 
Fourier mode $k_{_{C}}$. The density contrast evolves
proportional to the growing mode D(t), which is given in
\cite{pee80}.
Three of these linear inhomogeneities will be considered,
they correspond to 
$R_{_{C}}=250 \ Mpc$, $R_{_{C}}=50 \ Mpc$, and
$R_{_{C}}=12.5 \ Mpc \sim 8h^{-1} \ Mpc$,
and they
are denoted $L_{_{I}}$, $L_{_{II}}$, and $L_{_{III}}$,
respectively.
The derivative of the gravitational potential with respect 
to the physical radius $r$ is
\begin{equation}
\frac {d \phi } {dr} =\frac {1}{6} (\rho - \rho_{m0}) r
\label{dp3}
\end{equation}
for $r \leq aR_{_{C}}$, and
\begin{equation}
\frac {d \phi } {dr} = \frac {a^{2} R_{_{C}}^{2}} {r^{2}}
\frac {d \phi } {dr} (aR_{_{C}})
\label{dp4}
\end{equation}
for $r>aR_{_{C}}$.

\section{\label{sec3}Angular scales, boxes and resolution}

Recently, various groups have used simulations of structure
formation to estimate the effects of lensing on the CMB.
An important problem is that there are many scales producing 
lensing and that, according to White \& Hu \cite{whi01},
{\em simulating the full range of scales implied is currently
a practical impossibility}. 
These authors proposed the tiling method
to circumvent this problem and also the problems with
periodicity (see Sec. I); this method employs many independent
simulations with different boxes and resolutions to tile 
the photon trajectories. 
White \& Hu \cite{whi01} explained the advantages 
of their method 
with respect to the most traditional one based on plane projections,
which has been extensively used in the literature (see \cite{jai00}
and references cited therein).
The method proposed here does not require many independent simulations 
--but only one-- and, consequently, it is simpler and numerically
faster than the tiling one. Furthermore, it does not use plane projections 
whose potential problems were pointed out in \cite{whi01}.
In this section, we are concerned with the scales relevant
for CMB lensing, whereas periodicty is considered 
in next sections. 
 
We are interested in the deviations with respect to 
Gaussianity produced by lensing \cite{jai00}.
These deviations can be measured by 
angular correlations corresponding to three and more
directions.
Given an angular scale, 
the correlations are produced by a set of linear and nonlinear
cosmological structures and, a detailed study of the 
type of structures contributing to these correlations 
is necessary in order to design our N-body simulations,
which should include --in the same box-- all the structures
responsible for the effects we are looking for.
In order to estimate scales, we can
consider the evolving linear inhomogeneities
$L_{_{I}}$, $L_{_{II}}$, and $L_{_{III}}$,
and the nonevolving NFW density profiles 
$C_{_{I}}$, 
$C_{_{II}}$, and $C_{_{III}}$. Cluster evolution 
would not affect our qualitative estimates for small
redshifts $z<5$, although it could be crucial in other
contexts. In short, our study is based on six appropriate structures, 
three linear ones and three galaxy clusters.

\begin{figure*}
\includegraphics[width=16cm]{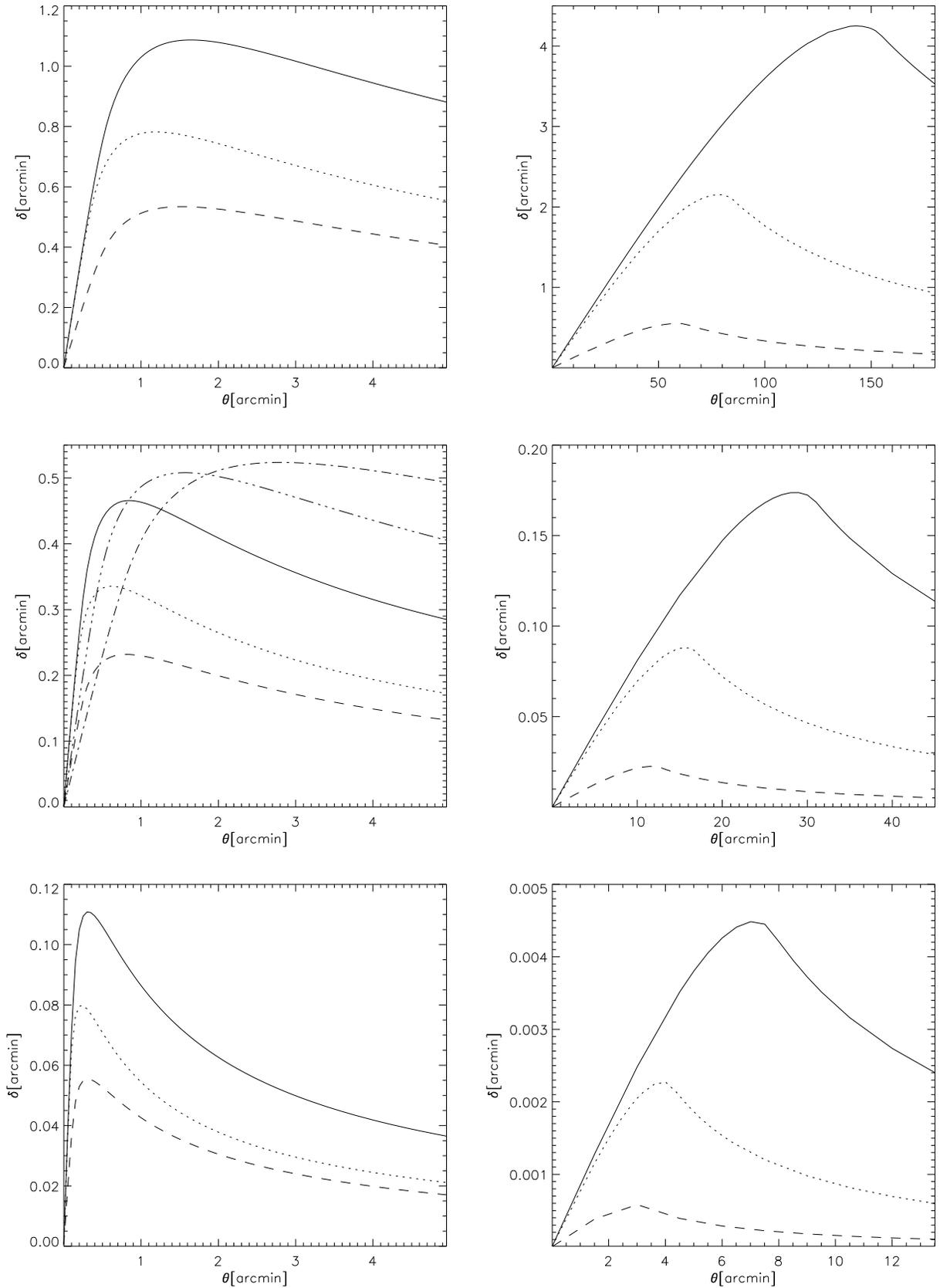}
\caption{\label{fig1}Lens deviation $\delta $ as a function of
the angle $\theta $ between the line of sight and the direction 
pointing towards the center of a spherical inhomogeneity.
Top, middle, and bottom right (left) panels correspond to the 
linear structures $L_{_{I}}$, 
$L_{_{II}}$, and $L_{_{III}}$ (clusters $C_{_{I}}$, 
$C_{_{II}}$, and $C_{_{III}}$) defined in the text.
In right (left) panel, solid, dotted, and dashed lines
correspond to redshifts 2, 10, and 100 (0.5, 2, and 5),
respectively. In the middle left panel, 
dotted-dashed (triple-dotted-dashed) line shows results for 
redshift 0.2 (0.1).} 
\end{figure*}

First of all, Eq. (\ref{devi}) and the potentials 
given in Sec. II have been used to calculate the lens deviations
produced by the six selected structures. Each linear
structure is placed 
at redshifts 2, 10 and 100, whereas each cluster is located
at redshifts 0.5, 2 and 5 (cluster $C_{II}$ is also placed at $z=0.2$ and
$z=0.1$). 
The deviation $\delta $ is calculated 
for each observation direction, which is characterized by
the angle $\theta $ formed by the line of sight and
the direction pointing towards the structure center. 
Results are presented
in Fig. \ref{fig1}, where we see that: 
(i) the deviations grow as the redshift decreases, and the 
growing is small for $z \leqslant 0.5$ 
(compare the curves of the left middle
panel of Fig. \ref{fig1} corresponding to redshifts 0.5, 0.2 and 0.1),  
(ii) the deviations increase as the inhomogeneity size
increases (this size is fixed by 
$M_{200}$ for clusters and by $R_{_{C}}$ for linear
structures), (iii) small linear structures
($L_{_{III}}$) produces very small deviations 
whatever their location may be, (iv) 
small clusters (or groups, $C_{_{III}}$) produce a maximum 
deviation which is about $10$ \% of the maximum deviation 
produced by cluster $C_{_{I}}$ (located at the same redshift), 
and (v) the most important 
deviations
are produced by massive clusters and by very 
extended linear inhomogeneities.

\begin{figure*}
\includegraphics[width=16cm]{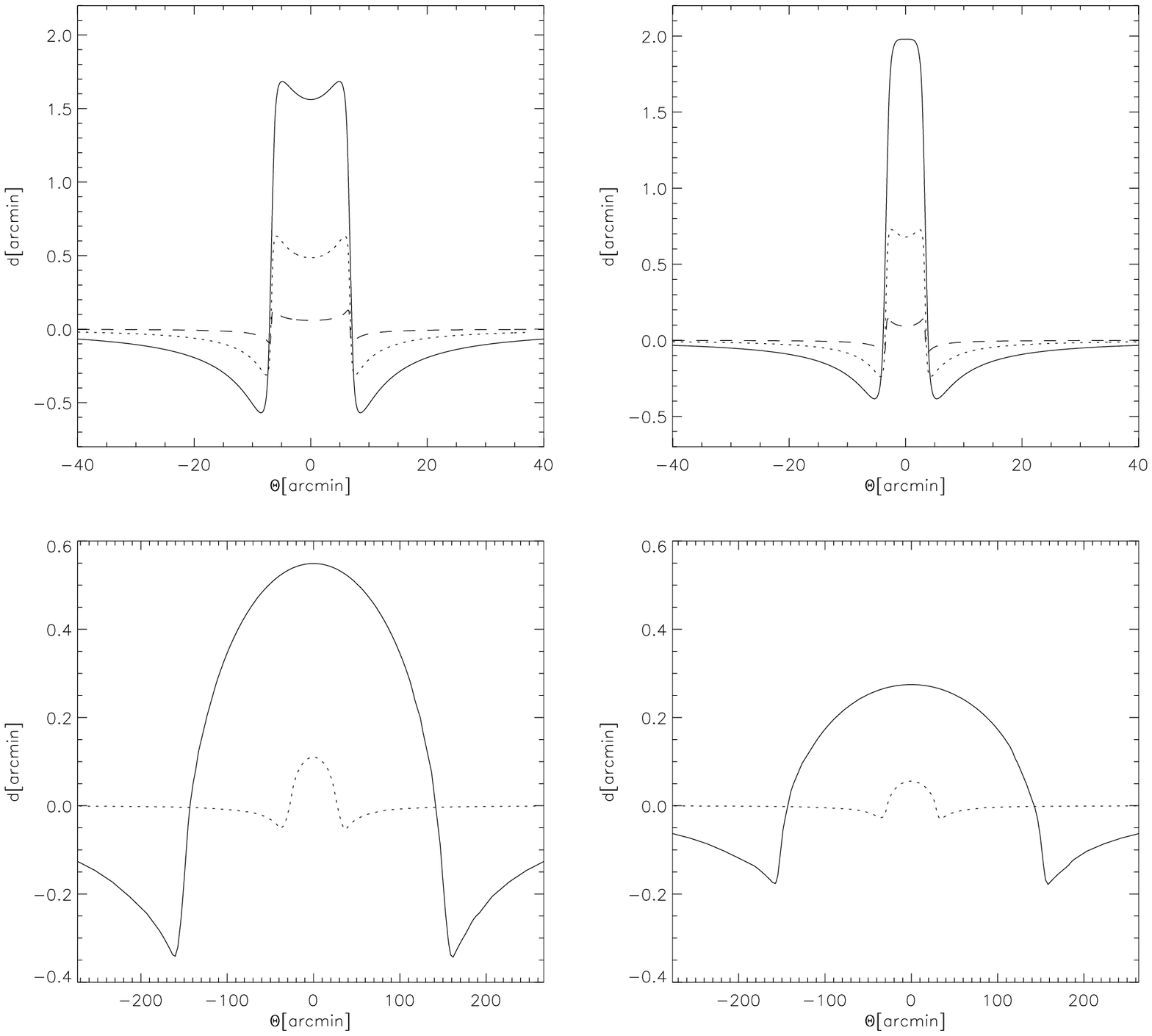}
\caption{\label{fig2}Lens deformation, $d$, 
of the angle $\alpha $ formed by two directions pointing 
towards the
same diameter of a spherical structure vs. the
angle $\Theta $ defined in the text. This angle fixes the 
relative position of these directions with respect 
to the line of sight of the symmetry center.
Solid lines of the top (bottom) panels
correspond to cluster $C_{_{I}}$ at $z=0.5$ 
(the $L_{_{I}}$ inhomogeneity at $z=2$),
whereas dotted lines correspond to the
same redshifts but different structures 
$C_{_{II}}$ and $L_{_{II}}$ in all the panels.
Dashed lines of the top panels are associated to cluster $C_{_{III}} $
at $z=0.5$.
Left (right) panels give deviations 
for $\alpha = 13.5^{\prime}$ ($\alpha = 6.75^{\prime}$).}
\end{figure*}

We are interested in the angular correlations 
of some CMB maps. Given a map of the variable $\zeta $ and $m$ directions, 
these correlations are defined as follows:
\begin{equation}
C_{m} = \langle \zeta (\vec {n}_{1}) \zeta (\vec {n}_{2}) 
\cdots \zeta (\vec {n}_{m}) \rangle \ ,
\label{corr}
\end{equation}
where the average is over many realizations
of the CMB sky. Some of these averages are estimated
below for two, three and four directions in maps of both
primary anisotropy and lens deformations.
For $m=4$, the chosen directions draw the vertices 
of a square on the last
scattering surface with 
$\vec {n}_{1} \cdot \vec {n}_{2} = \cos \alpha $, 
and for $m=3$, they point towards the vertices of 
an isosceles rectangle
triangle with $\vec {n}_{1} \cdot \vec {n}_{2} = \cos \alpha $
and $\vec {n}_{1} \cdot \vec {n}_{3} = \cos \alpha $.
The relevance of the deviations presented in the panels of
Fig. \ref{fig1} depends on the angular scale, $\alpha$, chosen 
for correlation computations.
Suppose the scale $\alpha $ ($\ell = \pi / \alpha $)
and $m=2$. 
From the primary temperatures corresponding to  
many pairs of directions 
($\vec {n}_{1}$, $\vec {n}_{2}$) forming angle $\alpha $, the average
$\langle \Delta_{_{P}} (\vec {n_{1}}) 
\Delta_{_{P}} (\vec {n_{2}}) \rangle$
can be calculated and the result measures  
correlations in the absence of lensing.
According to  Eq. (\ref{defor}), after lensing, the correlations 
are 
$\langle \Delta (\vec {n_{1}}) 
\Delta (\vec {n_{2}}) \rangle =
\langle \Delta_{_{P}} (\vec {n_{01}}) 
\Delta_{_{P}} (\vec {n_{02}}) \rangle$ and,
consequently, 
the correlations would change --due to lensing-- if, 
as a result of deviations, the directions
$\vec {n}_{01}$ and $\vec {n}_{02}$
form an angle $\alpha_{0}$ different
enough from $\alpha$; therefore, in order to see if
a certain spherical inhomogeneity 
located at redshift $z$ can contribute to 
the correlation at the angular scale $\alpha$, 
we may consider two directions pointing towards
$R$ and $S$, two points of a certain diameter 
of the spherical structure;
the first (second) of these directions forms angle $\theta $ 
($\theta + \alpha$ ) with the 
direction of the symmetry center, and the angle 
$\Theta = \theta + \alpha /2$ corresponds to the middle
point of the segment $RS$; then, 
the                    
differences $d(\Theta) = \delta(\theta +\alpha) - \delta(\theta)$
measure the deformations of the angle formed by the two chosen directions. 
The top panels of Fig. \ref{fig2} give these 
differences for 
the structures $C_{_{I}}$, $C_{_{II}}$, and $C_{_{III}}$ (at z=0.5), and
the bottom panels for $L_{_{I}}$ and $L_{_{II}}$ (at z=2).
The correlation scale is $\alpha = 13.5^{\prime}$ 
($\ell = 800$) in the left panels and $\alpha = 6.75^{\prime}$ 
($\ell = 1600$) in the right ones. From the analysis of this Figure it 
follows that: (a)
big close clusters of type $C_{_{I}}$ produce the largest deformations
reaching the maximum value $\sim 1.68^{\prime} $ 
($\sim 1.98^{\prime} $) for $\alpha = 13.5^{\prime}$
($\alpha = 6.75^{\prime}$),
(b) for $\alpha = 13.5^{\prime}$, inhomogeneities like 
$C_{_{II}}$, $C_{_{III}}$,
$L_{_{I}}$, and $L_{_{II}}$ produce maximum 
deformations which are about $37.5 \%$, $7.7 \%$, $33 \%$ and
$ 6.8 \% $ of the maximum ones corresponding to 
$C_{_{I}}$ clusters ($\sim 1.68^{\prime} $) and, (c) 
for $\alpha = 6.75^{\prime}$, the maximum deformations produced
by $C_{_{II}}$, $C_{_{III}}$,
$L_{_{I}}$, and $L_{_{II}}$ structures are about 
$36.9 \%$, $7.1 \%$, $13.7 \%$ and
$2.9 \% $ of the maximum deformation 
appearing in case $C_{_{I}}$ ($\sim 1.98^{\prime} $).
Points (b) and (c) lead to the conclusion that
the deformations produced by big linear structures
are more important for $\alpha = 13.5^{\prime}$, 
but they are not negligible in 
the case $\alpha = 6.75^{\prime}$. 
The deformations corresponding to
structures like $L_{_{III}}$ are too small
and they have not been presented in Fig. \ref{fig2}.

From the above discussion, it follows that
--for the scales under consideration--
the most relevant structures are clusters 
of type $C_{_{I}}$ and $C_{_{II}}$ at low redshifts, 
and big linear inhomogeneities
with $R_{_{C}} > 50 \ Mpc$ located at small redshifts.
This
conclusion is important. It tell us that simulations
which do not resolve galaxy groups and small clusters can
lead to good estimations, and
also that the box size can be chosen to prevent 
the existence of extended linear inhomogeneities
contributing to lensing. Boxes with 
sizes of $128 \ Mpc$ should be appropriated, sizes
of $256 \ Mpc$ could be acceptable (although a part of the 
lensing could be 
due to linear structures with $R_{_{C}} > 50 \ Mpc$, see Sec. V) and, 
finally,
boxes of $512 \ Mpc$ and greater could contain an extended part of 
a $L_{_{I}}$ structure giving important contributions
to lensing. The idea is that
the lens effect produced by big linear inhomogeneities can 
be studied without simulations, see references in \cite{jai00},
whereas the effect of any other significant structure 
($C_{_{I}}$ and $C_{_{II}}$ clusters) can be included 
in simulations for appropriate boxes
(between $128 \ Mpc$ and $256 \ Mpc$ size). 
Let us now analyze if results from Figs. \ref{fig1} and 
\ref{fig2} are robust against 
variations of the core radius $r_{c}$.

\begin{figure}
\includegraphics{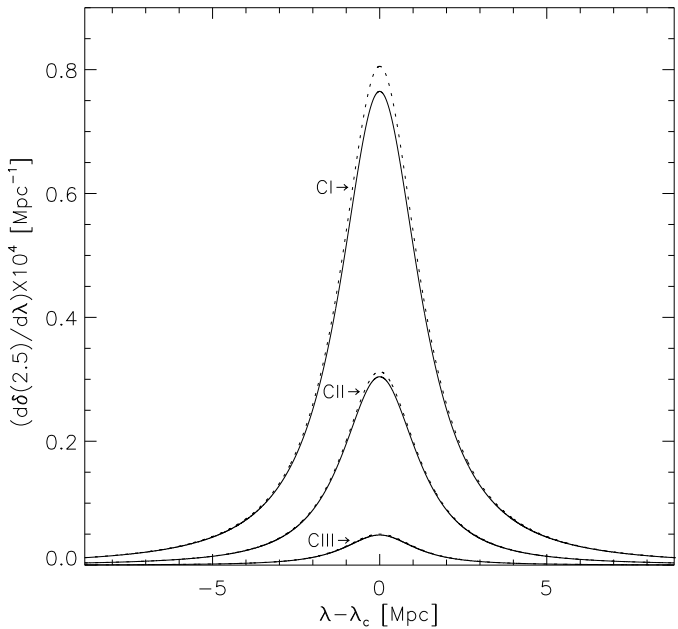}
\caption{\label{fig3}Function 
$2 W(\lambda) \vec {\nabla}_{\bot } \phi \times 10^{4} $
--involved in Eq. (\ref{devi})-- 
v.s. the comoving distance $\lambda - \lambda_{c} $ to the cluster center 
for the observation angle
$\theta = 2.5^{\prime }$.
All clusters have been located at $z = 0.5$. Each
pair of associated solid and dotted lines corresponds to the
cluster indicated inside the panel. Solid (dotted) line 
has been obtained from the greatest (smallest) values 
assigned to the core radius in the text. 
\vspace{2 cm}}
\end{figure}

Imagine one of the above three clusters
located at redshift z. 
It produces a significant deviation of the CMB photons
if and only if these photons   
cross the region where the function to be integrated 
in Eq. (\ref{devi})
(proportional to $\vec {\nabla}_{\bot } \phi$) is not negligible;
hence, if the dependence of this function on the 
core radius is weak, lens deviations (integral in Eq. (\ref{devi}))
also depend weakly on $r_{c}$, and previous results
based on Figs. \ref{fig1} and \ref{fig2} are robust.
In order to study the mentioned function for
clusters 
$C_{_{I}}$, $C_{_{II}}$, and $C_{_{III}}$, these 
structures are located at 
redshift 0.5 (where cluster evolution is not expected 
to be important) and, then,  
the function $d \delta(\theta) / d \lambda = 
2 W(\lambda) \vec {\nabla}_{\bot } \phi$ involved
in Eq. (\ref{devi}) is calculated at points of the 
null geodesic corresponding to the
angle $\theta = 2.5^{\prime}$. Results are   
presented in Fig. \ref{fig3}, where 
function $d \delta(2.5) / d \lambda$ is given 
(with arbitrary normalization) in terms of the comoving
distance to the cluster center $\lambda - \lambda_{c} $,
where $\lambda $ is the comoving distance from the observer 
to an arbitrary point of the null geodesic,
and $\lambda_{c} $ is the $\lambda $ value corresponding 
to the cluster center.
Continuous lines display the results obtained for 
clusters $C_{_{I}}$, $C_{_{II}}$, and $C_{_{III}}$ with
the radius cores we have previously fixed, whereas
the associated dotted lines correspond to the same
clusters with smaller cores, the new core radius 
being $0.1 \ Mpc$, $0.05 \ Mpc$, and $0.025 \ Mpc$
in cases $C_{_{I}}$, $C_{_{II}}$, and $C_{_{III}}$,
respectively (the initially chosen radius have been 
reduced by a factor $1/2 $).

The continuous 
and dotted associated lines of Fig. \ref{fig3}
are very similar for any cluster, which 
implies 
that, for $\theta = 2.5^{\prime} $, the lens deviation 
$\delta $ has a very weak dependence on the core radius 
$r_{c} $.
The direction $\theta = 2.5^{\prime} $ has been 
arbitrarily chosen in the $\theta $ interval where
the lens deformations are relevant (see
Fig. \ref{fig1}); nevertheless, other angles covering this 
interval have been also studied.
For example, we have investigated the directions corresponding 
to the maximum 
deviation in each of the cases reported in Fig. \ref{fig1},
and lens deviations also depends weakly on the
core radius. The same occurs for any $\theta$ value,
except for 
very small angles, which are not relevant for the 
angular scales under consideration (see below).
In Fig. \ref{fig3}, we can also see that 
the region where
$\vec {\nabla}_{\bot } \phi$ is significantly contributing
to the integral of Eq. (\ref{devi}) has a comoving size of 
various Megaparsecs for any cluster of interest.
It has been checked that the same conclusion
is also valid for any relevant $\theta $ values.

A scale for the computation of 
angular correlations has been chosen. In order to do that,
the CMB angular power spectrum ($C_{\ell}$ coefficients) 
due to lensing by nonlinear structures has been estimated
using CMBFAST. Results lead to the conclusion that 
$C_{2}(\alpha)$ correlations from this lensing
are important for $\alpha \leq 13.5^{\prime}$ ($\ell \geq 800$);
hence we must work with some scale $\ell > 800$. The scale
$\ell=1600 $ ($\alpha = 6.75^{\prime}$) has been 
chosen to calculate correlations 
because this scale has the following
properties: 
(a) for $\ell = 1600$, the effect of linear inhomogeneities --which should be 
reduced as much as possible--  is smaller
than that of the case $\l = 800 $
(compare the bottom panels of Fig. \ref{fig2}),
(b) according to CMBFAST estimations, 
lensing from nonlinear structures gives a $C_{1600} $ coefficient which is
smaller than the maximum $C_{\ell} $ only by a factor $\sim 0.6 $,
and (c) the value $\ell = 1600 $ 
is in the range of the multipoles to be observed by PLANCK
(the most sensitive projected experiment for CMB anisotropy 
measurements).

According to previous comments, a box of $128 \ Mpc$ have been 
assumed as the basic one and, then N-body simulations have been performed 
inside this box with a standard PM code \cite{hoc88}, which 
was used, tested, and described in detail in 
\cite{qui98}. The
resolution has been succesively increased 
starting from a very poor one of $2 \ Mpc $.
Of course, such a low resolution leads to non-linear structures
whose amplitudes (sizes) are smaller (larger) than 
those of the true clusters. These simulations only traces
a good spatial distribution of extended structures with total masses 
as those of the clusters (which would become clusters 
for appropriate resolutions). 
Resolutions of $1 \ Mpc $ and $0.5 \ Mpc $ have been also
considered,
but no higher resolutions have been used by the reasons given 
in Secs. V and VI.

\section{\label{sec4}Departures from Gaussianity}

Suppose an observer which is 
located at a certain point with comoving 
spatial coordinates $x^{i}_{_{P}}$. The main question is: which is the
deviation, $\vec{\delta }(\vec{x}_{_{P}},\vec{n})$, 
observed from
point $\vec{x}_{_{P}}$ in the $\vec{n}$ direction?

If $\phi_{\vec {k}} (t)$ is the FT
of the potential $\phi $, namely,
\begin{equation} 
\phi(\vec{x},t) = \frac {1} {(2 \pi)^{3/2}}
\int d^{3}k e^{-i \vec{k} \cdot \vec{x}}
\phi_{\vec {k}} (t)   \ ,
\end{equation}
and $\delta_{\vec{k}}$ is the FT of the density contrast 
$(\rho - \rho_{m0})/ \rho_{m0} $,    
Equation (\ref{cden}) leads to the following relation in
Fourier space: $\phi_{\vec {k}} = B \delta_{\vec{k}} / k^{2}$,
where $B = - \rho_{m0} / a$.

Using elemental Fourier algebra and Eqs.
 (\ref{devi}),
and (\ref{ng}), the following basic equations
--giving the required deviation-- are
easily obtained:
\begin{equation}
\vec{\delta }(\vec{x}_{_{P}},\vec{n}) =\frac {2i}{(2 \pi)^{3/2}}
\int  \frac {\vec{k}_{\bot }} {k^{2}}
F_{\vec {k}}(\vec {n}) e^{-i \vec {k} \vec {x}_{_{P}}} d^{3} k  \ ,
\label{funda1}
\end{equation}
where
\begin{equation}
F_{\vec {k}}(\vec {n}) = \int_{\lambda_{_{L}}}^{\lambda_{0}}
W(\lambda ) B(\lambda ) e^{-i \lambda \vec {k} \vec {n}}
\delta_{\vec {k}}(\lambda ) d \lambda  \ ,
\label{funda2}
\end{equation}
and $\vec{k}_{\bot} = \vec {k} - (\vec {n} \cdot \vec {k})
\vec {n}$.
  
According to Eq. (\ref{funda1}),
each component of $\vec{\delta }(\vec{x}_{_{P}},\vec{n})$ is
the FT --extended to all the space-- of a component of
vector $\vec{k}_{\bot }
F_{\vec {k}}(\vec {n}) / k^{2}$. The integral of the r.h.s. of 
Eq. (\ref{funda2}) must be estimated to get the
function $F_{\vec {k}}(\vec {n})$ involved in the FT. 
The integration variable  
$\lambda $ can be seen 
as a generalized time coordinate. Equations (\ref{funda1}) and
(\ref{funda2}) are general and they describe how the photons deviate 
--following
null geodesics-- in a realization of the full universe; 
nevertheless,
such a realization is not available in practice,
and we are constrained to use these equations 
in a fictitious periodic universe, let us now discuss this fact
and its consequences in detail.

Structure evolution is simulated with a certain N-body
code (see Sec. III), 
which involves a box and a certain resolution (a network 
in position space).
This code has a certain time step (or $\lambda$ step) and, 
consequently,
it gives data at a set of time values; among
these data, we are particularly interested in 
the function $\delta_{\vec {k}}(\lambda )$ 
which appears in Eq. (\ref{funda2}). This function is 
easily evaluated 
at each node, $\vec{k}$, in Fourier space, and at each time 
step $t_{i}$, but not at arbitrary times. The question is:
Can we calculate the integral (\ref{funda2})
using only the data corresponding to the time
discretisation of the N-body code? Fortunately, the answer
is positive. We have studied 
the function $\delta_{\vec {k}}(\lambda )$ 
for many $\vec {k}$ values and, thus, we have verified that  
the term $I(\vec {k}, \lambda ) =
W(\lambda ) B(\lambda ) \delta_{\vec {k}}(\lambda )$
--which appears in Eq. (\ref{funda2})-- 
can be very well 
approximated by a
straight line between $\lambda_{i}$ and $\lambda_{i+1}$, namely,
between
two successive time steps of the N-body simulation.
The equation of this line is of the form:
\begin{equation}
I_{i}(\vec {k}, \lambda ) = A_{i}(\vec {k}) \lambda + 
B_{i}(\vec {k})
\label{interpol}
\end{equation}
where quantities $A_{i}$ and $B_{i}$ are obtained from
$\delta_{\vec {k}}(\lambda_{i} )$  and
$\delta_{\vec {k}}(\lambda_{i+1} )$; namely, from quantities given 
by the N-body in two successive steps. 
If Eq. (\ref{interpol}) is substituted into 
Eq. (\ref{funda2}), the integral of the r.h.s. can be
analytically calculated in the interval ($t_{i}, t_{i+1}$). 
The addition of these integrals for all the N-body 
intervals gives function $F_{\vec {k}}(\vec {n})$.
After this integration, the FT (\ref{funda1}) can
be performed and, thus, {\em the deviation field in the $\vec {n}$
direction is simultaneously calculated for many observers located
at the nodes $\vec{x}_{_{P}}$
of the spatial Fourier box}. The importance of this fact is discussed
below.

\begin{figure*}
\includegraphics[width=16cm]{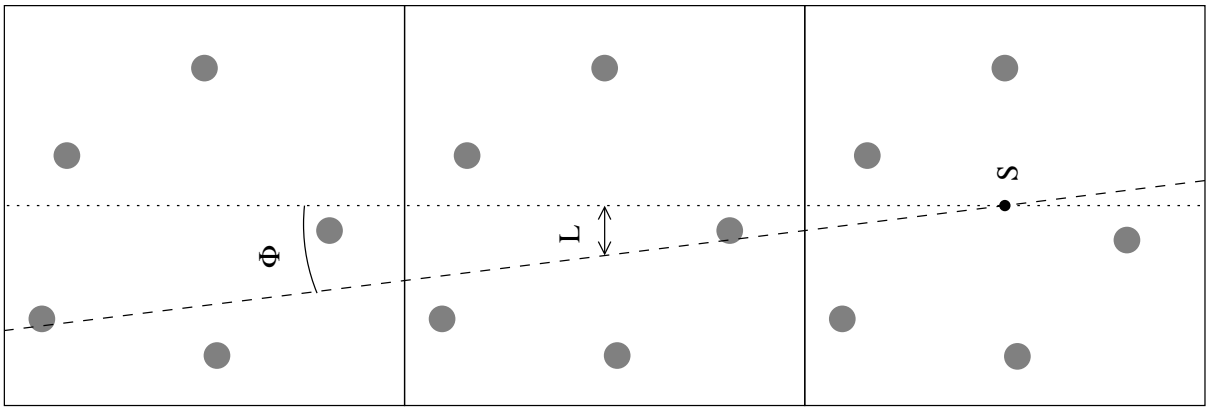}
\caption{\label{fig4}Sketch of a 2D universe filled by 
squares which contains clusters (circles) and photon patches (lines). 
Dotted (dashed) line corresponds to a direction 
parallel (non parallel) to the square edges.}
\end{figure*}

     Discretisation implies that the CMB photons do not move 
in a true realization of the universe, but in a periodic
universe, which is, at each time,  
an ensemble of boxes identical to that of
the N-body simulation. This periodicity should lead to errors
in our estimations. In order to understand the problem and 
its solution, let us establish a two-dimensional (2D) analogy.
The squares of Fig. \ref{fig4} (boxes in the 3D case) have a size of
$128 \ Mpc$. These squares contain small circles (clusters), and 
let us imagine an observer, S, at the center of the right square.
If a photon propagates along the pointed line
of Fig. \ref{fig4}, it crosses all the squares in the same way, namely,
it passes at the same distances from the same
clusters and undergoes the same systematic repeated 
deviations. Only evolution 
induces small deformations from square to square, but 
these deformations would be only
relevant after a large enough number of 
squares have been crossed; hence, 
a certain wrong accumulative effect should appear. However,
if the photon moves following the dashed line of Fig. \ref{fig4},
the situation is very different because the photon enters 
each square around a very different place and, then, it can be 
affected by distinct independent clusters.
The key point is the selection of the
angle $\Phi $ and its associate
distance $L$. Two facts are relevant to make this choice:
(i) clusters significantly deviate the CMB photons when their
impact parameter is smaller than 
a few Megaparsecs (see Fig. \ref{fig3}), and (ii) correlations
in the cluster distribution (superclusters) extend
near $10 \ Mpc$; hence, if we choose $L$ between $15 \ Mpc$
and $20 \ Mpc$, photons 
move from square to square, but they are always influenced
by different almost uncorrelated clusters and, consequently, the
effect of periodicity must be minimum. All these
considerations are easily generalized to the 3D case,
in which there are two angles defining the 
line of sight. Hereafter, these angles are identified 
with the spherical coordinates  
$\theta $ and $\phi $ (with respect to the Cartesian axis
defined by the edges of the box).
These  angles can be easily chosen to maximize the
length of the path where the photons undergo the
action of distinct uncorrelated clusters; namely, to minimize 
the effects of periodicity. In the 3D case, with boxes
of $128 \ Mpc$, it has been found that 
for $\theta = 10.75^{\circ}$ and $\phi = 41.19^{\circ}$,
the photons would cross 56 boxes before arriving
to the initial region of the box. The CMB photons moving
along this direction (hereafter denoted $\vec {N}_{0}$ and called 
the preferred direction)
do not feel periodicity during their
travel from $r \simeq 7400 \ Mpc$ to $r=0$, namely, from 
$z=3.62$ to $z=0$ and, taking into account that the lens
effect we are evaluating
is produced at these low redshifts by galaxy clusters, 
it follows that, for direction $\vec {N}_{0}$ and close 
ones, periodicity must be rather irrelevant (see Sec. V 
for more discussion).
Furthermore, 
the photon enters in two successive boxes through points that
--when placed in a unique box--
are located at a distance $L = 15.9 \ Mpc$. 
For the chosen values of
$\theta $ and $\phi $ and a box of $256 \ Mpc$, we have 
$L = 31.8 \ Mpc$, and photons does not feel periodicity
at all from decoupling to present time.

On the angular scales under consideration,
the cluster distribution producing lensing is statistically 
independent on the structure distribution
causing the primary anisotropy;
in fact, for $\alpha$ values between $6.75^{\prime}$ and 
$13.5^{\prime}$ ($1600 \geq \ell \geq 800$), primary anisotropy 
is due to temperature fluctuations previous to decoupling, which
appeared as a result of the coupling between 
matter an radiation in the presence of energy density fluctuations;
hence, the scales of temperature fluctuations and those of
energy perturbations coincide.  Inhomogeneities with a comoving
size of $41 \ Mpc$ ($20.5 \ Mpc$) located at $z=1100$ 
subtend an angle of $13.5^{\prime }$ ($6.75^{\prime }$); hence,
taking into account that --in our $\Lambda$CDM model--
the comoving scale
of an Abell cluster with $10^{15} M_{\odot }$ is $25.6 \ Mpc$,
we can conclude that, for $800 \leq \ell \leq 1600$, the
primary anisotropy is produced by
comoving scales similar to those of clusters. 
These small structures located close to our last scattering 
surface ($z \sim 1100$) and the clusters producing 
lensing (small $z$ values)
are very far, different, and independent structure
distributions; therefore,
we can consider that the distribution of clusters is 
producing deformations
on independent primary CMB maps. In order to calculate 
the averages of Eq. (\ref{corr}) (correlations), various 
full realizations 
of the cluster distribution should be considered
to produce deviation fields and, then,
each of these fields should be used to deform a large enough number of
independent primary maps. In practice (computational limitations), 
only a few pairs of directions 
(close to $\vec {N}_{0}$ to minimize periodicity
effects)
are studied for each 
cluster distribution (N-body simulation), and these directions should 
be used to deform different primary maps. Let us now describe
in detail the method implemented in this paper 
to calculate correlations.

For each N-body simulation, nine 
directions $\vec {n}_{i}$ are considered. These directions
are chosen in such a way that they depict the vertices of four 
neighboring squares with size $\alpha $ on the last scattering surface.
These squares form another 
bigger one
with size $2 \alpha $ with vector $\vec {N}_{0}$ pointing
towards its center. 
It is important that, for each of the chosen
directions, the deviations 
$\vec{\delta }(\vec{x}_{_{P}},\vec{n}_{i})$ are calculated 
for the $N^{3}$ observers, $\vec{x}_{_{P}}$,  
at the same time (see above). Let us discuss the importance
of this fact from the statistical point of view.
Put a $4.25^{\circ} \times 4.25^{\circ}$
map, $M_{_{P}}$, of primary anisotropy in such a way that,
from the box center (point $\vec{x}_{_{C}}$), the direction   
$\vec {N}_{0}$ points towards the map center $C$. Consider now
one of the observation direction $\vec{n}_{i}$.
From the box center, the chosen direction points towards a
certain point $C_{i}$; nevertheless, from another 
point of the box $\vec{x}_{_{Q}}$, direction 
$\vec{n}_{i}$ does
not point towards point $C_{i}$ but towards another
point $Q_{i}$ of the map $M_{_{P}}$. 
Two parallel lines --with direction $\vec{n}_{i}$--
starting from points 
$\vec{x}_{_{C}}$ and $\vec{x}_{_{Q}}$ of the box 
would intersect the map $M_{_{P}}$ in points $C_{i}$ and 
$Q_{i}$, respectively;
hence, the $Q_{i}$ coordinates in the map $M_{_{P}}$ can be 
calculated as follows: project the vector
$\vec{x}_{_{Q}} - \vec{x}_{_{C}}$ on the plane
normal to $\vec{n}_{i}$, translate this projection parallely 
from the hypersurface $t=0$ to the last scattering surface 
at $z=1100$, calculate the angle subtended by the
final vector (as it is seen by the observers) and, then, 
the angular coordinates
of point $Q_{i}$ in the map $M_{_{P}}$ 
can be trivially obtained. 
Of course, the
deviation, $\vec{\delta }(\vec{x}_{_{Q}},\vec{n}_{i})$, 
corresponding to the observer $\vec{x}_{_{Q}}$ in the
direction $\vec{n}_{i}$,
must be applied to point $Q_{i}$ in the map $M_{_{P}}$. 
Hence, for each direction $\vec{n}_{i}$, we have not 
a unique point in $M_{_{P}}$,
but many points, one for each 
$\vec{x}_{_{Q}}$ observer, and these points 
cover a certain region on map $M_{_{P}}$.  
In order to estimate the size of the covered region, we can consider 
a fixed direction, for example
$\vec {n} = (0,0,1)$, 
and a box of $128 \ Mpc$  ($256 \ Mpc$) per edge; thus,
given two observers $\vec{x}_{_{Q1}}$ and $\vec{x}_{_{Q2}}$ separated by 
a distance of $128 \ Mpc$ ($256 \ Mpc$),
the angular distance between points $Q_{1}$ and $Q_{2}$ 
in the map $M_{_{P}}$ appears to be 
$\Xi = 0.5^{\circ}$  ($\Xi = 1^{\circ}$), which means that, 
for $\alpha = 13.5^{\prime}$, 
a squared region with a size of
$43.5^{\prime}$  ($73.5^{\prime}$) is covered,
and for $\alpha = 6.75^{\prime}$, the size of the corresponding region
is $36.75^{\prime}$  ($66.75^{\prime}$); hence,
we see that, for boxes of $128 \ Mpc$ ($256 \ Mpc$) and
$\alpha=6.75^{\prime}$, the size of
this region is larger than $\alpha $ by a factor close to five
(ten) and,
consequently, 
the averages in the $\vec{x}_{_{Q}}$ observers 
--a novel aspect of our ray-tracing procedure--
play
a crucial statistical role. The greater the boxes and the smaller
the angular scales $\alpha $, the larger the above factor and the 
greater the statistical significance of the $\vec {x}_{_{Q}}$
averages.

On account of the above considerations, correlations are calculated 
as follows: (i) twelve different pairs of directions, sixteen triads, 
and four tetrads are defined using the above nine
directions, (ii) given an observer $\vec{x}_{_{Q}}$,
a map $M_{_{P}}$ (primary anisotropy), and a N-body
simulation (for $\vec {\delta }$ computation),
the deviations  $\vec{\delta }(\vec{x}_{_{Q}},\vec{n}_{i})$
and Eq. (\ref{defor})
(or Eq. (\ref{dmap})) are used to calculate the lens effect 
$\Delta [\vec {n}_{i} + \vec{\delta }(\vec{x}_{_{Q}},\vec{n}_{i})]
-\Delta [\vec {n}_{i}]$ (or $ \vec{\delta }(\vec{x}_{_{Q}},\vec{n}_{i})
\cdot [\partial \Delta_{_{P}} / \partial \vec{n}]_{\vec{n}_{i}} $) 
for each of the 
$\vec{n}_{i}$ directions and, then, the resulting data are
used to compute the averages (\ref{corr}) for the above
pairs ($m=2$), triads ($m=3$), and tetrads ($m=4$),
(iii) calculations in (ii) are repeated for each observer 
$\vec{x}_{_{Q}}$, and results are averaged again, (iv) 
all the process is repeated for a certain number 
of $M_{_{P}}$ maps and a new average is performed and,
(v) finally, the above calculations are repeated for
a certain number of N-body simulations, and the final averages are 
taken as our
estimations of the correlations (\ref{corr}). The question is: 
How many primary maps
and how many cluster realizations would be necessary to
calculate the required correlations? 

In order to answer this question, the following method
has been implemented: In a first step, the process (i)-(iv)
has been used to analyze pure primary maps in the
absence of lens deviations. The primary maps are not deformed 
and, consequently, the resulting averages 
should approach the true two, three, and four direction 
correlations of these maps, which are hereafter called 
primary correlations. These correlations 
can be estimated by
a direct analysis of the simulated primary
maps (see next section) and the results of such an analysis
can be compared with the correlations estimated by the process
(i)-(iv) in the absence of deviations.
Since no deviations have been considered, this study
measures the capability of our method 
(based on nine direction, $N^{3}$ observers, and various 
primary maps)
to create pairs,
triads and tetrads covering a significant part of
the primary maps; namely, allowing a good statistical
analysis of maps.
In the second and final step, the number
of primary maps suggested by our previous study 
(in the absence of deviations) is taken 
and, then,
deviations from more and more N-body simulations
are considered in order to compute the required correlations 
as it has been described above.
When the averages reach almost stable values, 
the process is stopped and no
more N-body realizations are performed.
Results are presented in next section.

\begin{figure*}
\includegraphics[width=16cm]{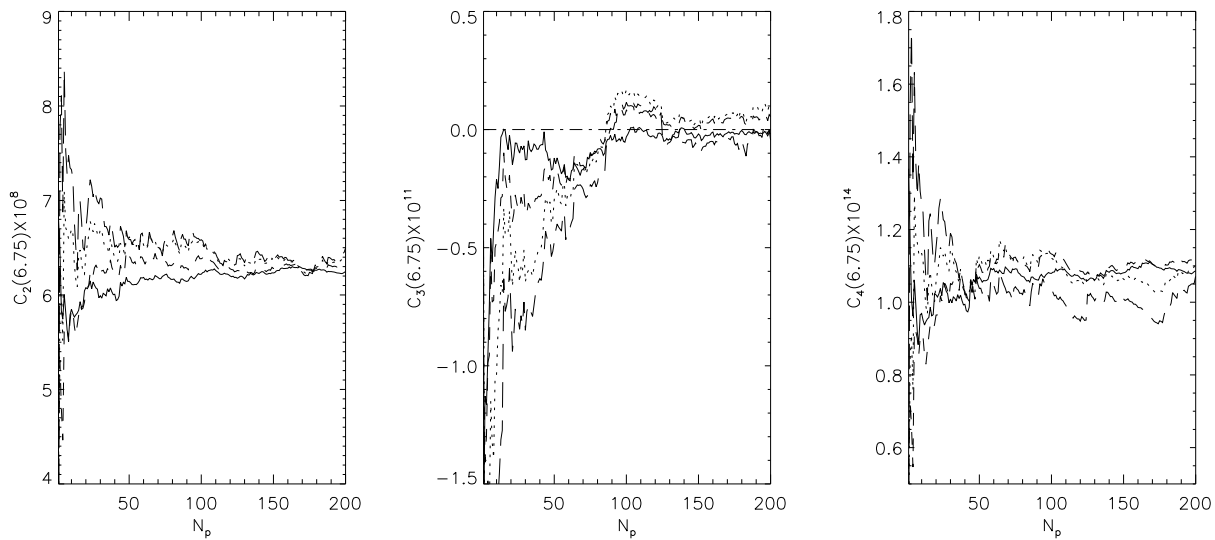}
\caption{\label{fig5}Correlations $C_{2}$, $C_{3}$, and $C_{4}$
extracted from simulated maps of the primary 
CMB anisotropy (for $\alpha = 6.75^{\prime}$),
versus the number $N_{p}$
of simulated maps used in the computations.
Left, central, and right panels show the
correlations $C_{2}$, $C_{3}$, and $C_{4}$, respectively.
Different curves correspond to distinct methods for correlation
estimates which are described in the text.}
\end{figure*}

\section{\label{sec5}Results}

First of all, the methods applied --in this paper-- to calculate 
correlations of the lens effect are tested. In order to do that,
these methods are used to analyze well known maps of primary 
anisotropy. The angular scale for correlation estimates is
$\alpha =6.75^{\prime }$.
A set of two hundred maps of primary anisotropy has been used.
Results are presented in Fig. \ref{fig5}, 
where variable $N_{p}$ (horizontal axis) 
is the number of primary maps used to
get the correlation value appearing in the vertical axis.
Left, central, and right panels
correspond to the two ($m=2$), three ($m=3$) and four ($m=4$)
direction correlations (\ref{corr}), respectively. 
In all the panels,          
the solid line shows the correlations obtained by means of an
exhaustive coverage of the $M_{_{P}}$ maps using many
pairs, triads and tetrads 
of directions. This is our best estimate of the 
required correlations; results from this method 
--which cannot be applied to analyze the lens component-- approach 
the theoretical values of
the correlations we have used to simulate the $M_{_{P}}$ maps. 
Other methods must give similar correlations to be acceptable.
As $N_{p}$ increases, the solid lines of the top and bottom panels 
seem to tend to a certain value, 
whereas the continuous line of the middle panel
seems to be compatible with a vanishing 
$C_{3}$ correlation.
The long dashed lines
display the correlations obtained using nine directions 
and only one observer (only nine directions in each map).
These lines slowly approaches the solid one as $N_{p}$ 
increases and the level of approximation 
seems to be good for $N_{p} = 200$. 
The dotted (dashed) lines 
give the correlations obtained with
nine directions and
$128^{3}$ ($256^{3}$) observers. As it  
follows from Fig. \ref{fig5},
dashed and dotted curves approach the solid ones faster than the 
long dashed lines and, consequently, the use of
many observers located in the nodes of a network 
(see Sec. IV) is statistically significant; 
the greater the box size, the faster the statistical approximation
to the solid lines. 
The best situation corresponds to 
$256^{3}$ observers in a $256 \ Mpc$ box. 
Two hundred primary maps 
suffice for a good enough estimate of $C_{2}(6.75)$ and
$C_{4}(6.75)$; particularly, when $N^{3}$ observers
are considered. 

After computing the correlations $C_{2}(6.75)$, $C_{3}(6.75)$,  
and $C_{4}(6.75)$ for lens deformations,
it would be worthwhile a comparison between the correlation level
of these deformations and that of primary maps.
The ratios 
$r_{2} = C_{2}(6.75)/ C_{2}(0)$,
$r_{3} = C_{3}(6.75)/ C_{2}^{3/2}(0)$, and
$r_{4} = C_{4}(6.75)/ C_{2}^{2}(0)$ 
are apropriate to
do this comparison; these quantities 
will be calculated for lens deformations and
primary maps and, then, results will be compared.
We begin with the case of primary maps. Using 
two hundred of these maps and exhaustive coverage, we have found:          
$C_{1} = \langle \Delta_{_{P}} (\vec {n}) \rangle =
-2.12 \times 10^{-6} $, 
$C_{2}(0) = \langle \Delta_{_{P}} (\vec {n}) \Delta_{_{P}} 
(\vec {n})\rangle = 7.29 \times 10^{-8}$, and 
$r_{1} = C_{1}/ C_{2}^{1/2}(0) = 7.85 \times 10^{-3}$. 
We know that 
there are many positive and
negative temperature contrasts --in the maps--
with absolute values close
to $C_{2}^{1/2}(0)$ and, consequently, 
only an strong cancellation of these contrasts
can explain the resulting value of the 
mean contrast $C_{1}(0)$, which is much smaller
than $C_{2}^{1/2}(0)$ (small ratio $r_{1} \sim 7.9 \times 10^{-3}$).
Strong cancellation leading to a small $r_{1}$ value
indicates that, as the number of 
maps increases, quantity  
$\langle \Delta_{_{P}} (\vec {n}) \rangle$
approaches either a small value or zero. Similarly, in order to calculate
the correlation
$C_{2}(6.75) = \langle \Delta_{_{P}} (\vec {n_{1}}) 
\Delta_{_{P}} (\vec {n_{2}}) \rangle$, positive and 
negative numbers with absolute values close to 
$C_{2}(0)$ are averaged. The resulting ratio 
$r_{2} \sim 0.85$ indicates that 
these numbers have not cancelled among them and
a nonvanishing $C_{2}(6.75)$
correlation exists. For $m=3$ (m=4), the relevant ratio is
$r_{3} \sim 8.8 \times 10^{-3}$ 
($r_{4} \sim 2$) and, this small (large) ratio $r_{3}$ ($r_{4}$)
strongly suggests a vanishing $C_{3}$ (significant $C_{4}$)
correlation.

Since the simulated primary maps are Gaussian,
the relations $C_{3} (6.75) = 0$ and 
$C_{4} (6.75) = 3 C_{2}^{2} (6.75)$ must be satisfied.
The correlation $C_{3}$ obtained by exhaustive coverage 
(solid lines of Fig. \ref{fig5}) is $C_{3} \simeq  -1.76 \times 10^{-13}$, and
it has been proved to be small
by computing $r_{3}$; furthermore, 
we have used the same method to get the correlations
$C_{2} \simeq  6.24 \times 10^{-8}$
and
$C_{4} \simeq  1.09 \times 10^{-14}$, which lead to the ratio 
$3 C_{2}^{2}/C_{4} \simeq 1.07$; hence,
we can say that the correlations extracted from the maps are compatible 
with Gaussianity, as it should be for the primary
maps we are analyzing. If we use other methods to analyze the 
$M_{p}$ maps; for example, the methods used to built up the dotted, 
dashed, and long dashed curves of Fig. \ref{fig5} (see above),  the 
extracted correlations appear to be similar to those reported above 
and so they are
compatible with Gaussianity; this means that
all the methods used to analyze maps work very well for
primary maps, and the same should occur in our applications 
to the analysis of lens deformations.

\begin{figure*}
\includegraphics[width=16cm]{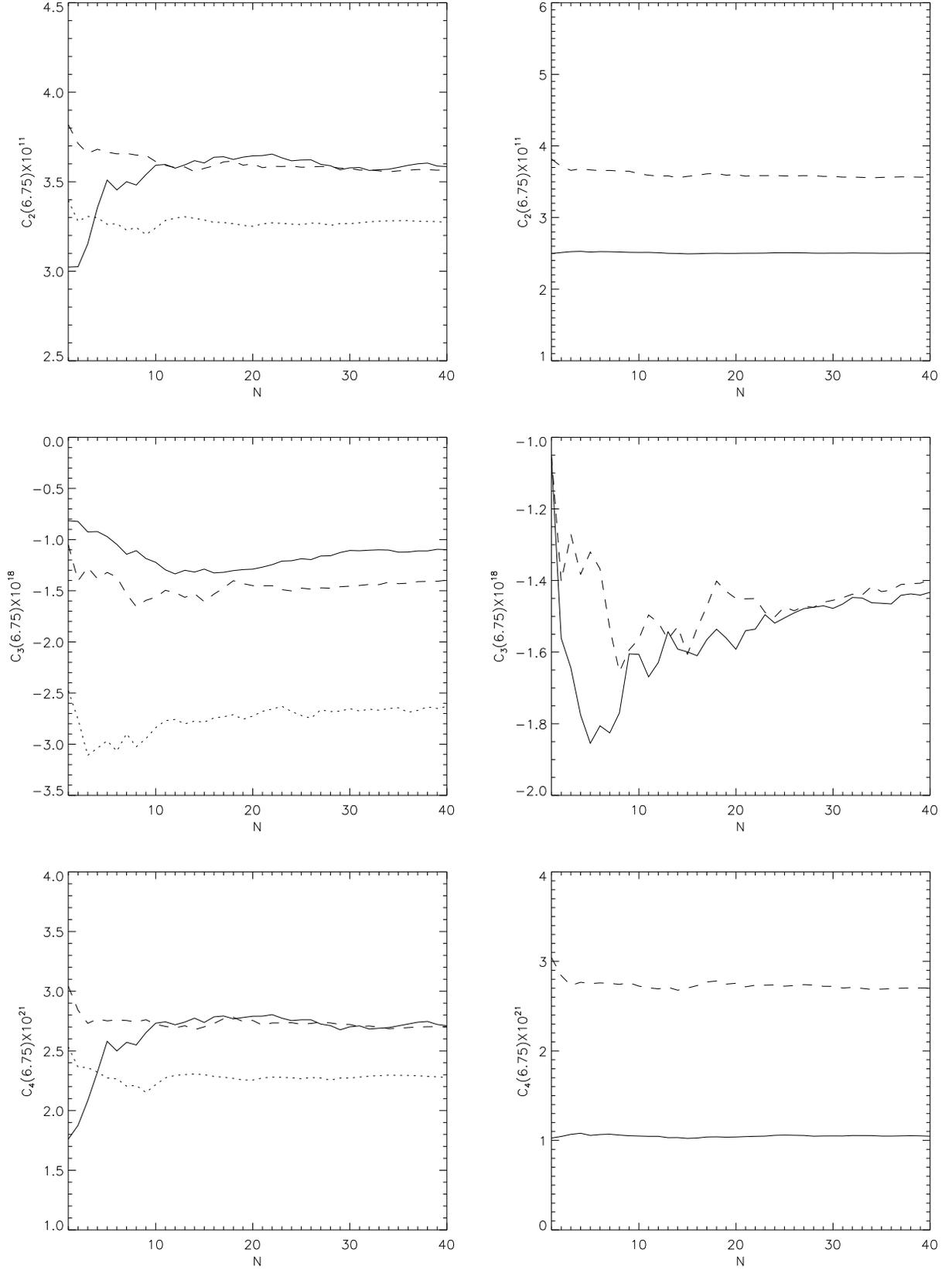}
\caption{\label{fig6}Correlations $C_{2}$, $C_{3}$, and $C_{4}$
extracted from simulated maps of the lens effect
(for $\alpha = 6.75^{\prime}$), versus     
the number $N$
of simulated maps used in the calculations.
Top, middle, and bottom panels show the
correlations $C_{2}$, $C_{3}$, and $C_{4}$, respectively.
solid, dashed, and dotted lines of the left panels corresponds
to resolutions of $0.5 \ Mpc$, $1 \ Mpc$, and $2 \ Mpc$,
respectively. Dashed (solid) lines of the right panels
show correlations for direction close to the preferred one (a
direction parallel to an edge of the simulation box).}  
\end{figure*}

\begin{table*}
\caption{\label{tbl-1}Correlations in simulated lens 
deformations of the CMB sky.}
\begin{ruledtabular}
\begin{tabular}{ccccccc} 
Case  &  Box Size  & Resolution  & Direction 
& $C_{2} \times 10^{11}$ & $C_{3} \times 10^{18}$  &
$C_{4} \times 10^{21}$ \\
&in Mpc & in Mpc& & & &\\ 
\hline  
1 &128 &0.5 &optimal &3.59 $\pm $ 0.12  &$-$1.10 $\pm $ 0.14 &2.71 $\pm 
$ 0.18\\
2 &128 &1.0 &optimal &3.57 $\pm $ 0.06  &$-$1.40 $\pm $ 0.18 &2.70 $\pm 
$ 0.10\\
3 &128 &2.0 &optimal &3.27 $\pm $ 0.06  &$-$2.63 $\pm $ 0.24 &2.28 $\pm 
$ 0.08\\
4 &128 &1.0 &parallel to an edge &2.50 $\pm $ 0.02  &$-$1.43 $\pm $ 0.16 
&1.05 $\pm $ 0.03\\
5 &256 &1.0 &optimal &3.08 $\pm $ 0.06  &$-$0.76 $\pm $ 0.06 &2.07 $\pm 
$ 0.08\\
\end{tabular}
\end{ruledtabular}
\end{table*}

Correlations of lens deformations have been calculated in various 
cases using distinct PM simulations;
in case 1, the resolution is
$0.5 \ Mpc$ (the best one), the box size is $128 \ Mpc$, directions are close
to the preferred one, and the angular scale is $\alpha = 6.75^{\prime}$. 
Correlations obtained in this case are 
presented in the 
first row of table~\ref{tbl-1}, and also in the continuous lines of the left
panels of Fig. \ref{fig6}. All the data of 
table~\ref{tbl-1} are mean correlations with 
$2 \sigma $ errors calculated from the correlations of 
forty simulations. These errors are statistical ones assigned 
in case 1, but there are other sources of uncertainty as periodicity and
simulation features, which are not included in 
the errors reported in table~\ref{tbl-1}. 
The lines mentioned above display the variation of the estimated
correlations as the number N of simulations (used to
perform averages) increases; 
for $N > 20$, correlations $C_{2}$ and $C_{4}$ seem to coverge
towards a certain value,
whereas the correlation $C_{3}$ slowly decreases in the
$N$-interval (20,40) 
(see continuous lines in the middle panels of Fig. \ref{fig6}, which 
are two different representations of the same function). 
We assign to $C_{2}$, $C_{3}$, and $C_{4}$ the values and errors given
in the first row of table~\ref{tbl-1}; 
the tabulated values of $C_{2}$ and $C_{4}$
are estimations of these correlations, whereas that of 
$C_{3}$ can be better considered as an upper limit 
due to the slow decreasing we have mentioned above. 
The values assigned to $C_{2}$, $C_{3}$, and $C_{4}$ have been used to
calculate the ratios $r_{2} \simeq 0.081 $, 
$r_{3} \simeq 1.17 \times 10^{-4}$, and $r_{4} \simeq 0.014$,
which are smaller than the corresponding ratios of the 
primary maps (see above). The small $r_{3} $ value indicates that lensing 
(by nonlinear structures) does
not introduce any appreciable $C_{3}$ correlation, whereas
the small values of $C_{2}$ and $C_{4}$ suggest small levels of
these correlations. Finally, the ratio  
$3 C_{2}^{2}/C_{4} \simeq 1.43$ confirms a deviation with respect
to Gaussianity. 

\begin{table}
\caption{\label{tbl-2}Comparing lens correlations for various 
pairs of estimations.}
\begin{ruledtabular}
\begin{tabular}{cccc}
Compared Cases  & $100 \Delta C_{2} / C_{2}$ & $100 \Delta C_{3} / C_{3}$ 
& $100 \Delta C_{4} / C_{4}$\\
\hline
1-2 &0.56 &21. &0.37\\ 
2-3 &8.91 &58. &16.\\
2-4 &30 &2.09  &61\\
2-5 &13.73 &46 &23\\  
\end{tabular}
\end{ruledtabular}
\end{table}

Since the method 
proposed in previous sections 
is based on  N-body simulations, results could depend on 
both resolution 
and box size. These dependences are now studied. 
In order to analyze the importance of resolution, 
cases 2 and 3 have been 
considered, they are identical to the case 1 described above except 
for resolution, which is $0.5 \ Mpc$ in case 1, 
$1. \ Mpc $ in case 2, 
and $2. \ Mpc $ in case 3. 
Results of case 2 (3) are displayed in the second (third)
row of table~\ref{tbl-1}, and also in the dashed (dotted) lines of the 
left panels of Fig. \ref{fig6}, where we see that
the correlations $C_{2}$, $C_{3}$, and $C_{4}$ of case 3
($ 2. \ Mpc$ resolution) are smaller than          
those of case 2 ($1. \ Mpc $) 
and case 1 ($0.5 \ Mpc $), which are very similar between them.
In order to quantify these comparisons between results
corresponding to pairs of cases, the 
relative variations $\Delta C_{m} / C_{m}$ for $m=2,3,4$
are obtained for the pairs 
1-2 and 2-3, and the 
comparation percentages $100 \Delta C_{m} / C_{m}$ for these pairs 
are given in the first and second rows
of table~\ref{tbl-2}, where we see that the comparation percentages
of the pair 1-2 are much smaller than those of 
2-3. These considerations 
strongly suggests that a $2. \ Mpc $ resolution
is poor, whereas a $1. \ Mpc $ resolution is good because 
a better one ($0.5 \ Mpc $) does not lead to significantly better results; 
particularly,
for the correlations $C_{2} $ ($100 \Delta C_{2} / C_{2} = 0.56$
for cases 1-2) 
and $C_{4} $ ($100 \Delta C_{4} / C_{4} = 0.37$ for cases 1-2).
Taking into account that cluster description can be 
improved using higher resolutions,
these results have two possible interpretations: (1)
the calculation of the lens correlations 
does not require a 
more detailed description of the clusters (for the
chosen angular scale $\alpha = 6.75^{\prime} $) and,
(2) results from resolutions of $1 \ Mpc $ and $0.5 \ Mpc $
are similar because of a lack of resolution in both cases,
but higher resolutions would lead to --different-- better results. 
We have not found any theoretical reason supporting one of 
these alternatives.
In this situation, the most effective procedure 
is the use of new simulations with higher 
resolutions, which should 
support one of the above possibilities
(see Sec. VI for more discussion).

After concluding that a resolution of $1. \ Mpc $ 
seems to be good (at least it is equivalent to a
better resolution of $0.5 \ Mpc $),
and before analyzing the importance of the box size,
a study of the relevance of periodicity is worthwhile.
In order to perform such a study, results of case 2
are compared with those of case 4. In both cases, the                                                 
same forty simulations are used,
but the propagation directions of the
CMB photons are different. In cases 2 and 4, directions form small angles
with the preferred direction and with
an edge of the simulation box, respectively.
Results of case 4 are displayed in the forth 
row of table~\ref{tbl-1}, and also in the solid line of the 
right panels of Fig. \ref{fig6} (where dashed lines correspond to
case 2). Comparing the solid and dashed lines of these
panels, one easily concludes that periodicity strongly (weakly) affects
the correlations $C_{2} $ and $C_{4} $ 
($C_{3} $). Accordingly,  
the comparation percentages of the correlations 
$C_{2} $ and $C_{4} $ for the pair of cases 
2-4 (third row of table~\ref{tbl-2})
are much greater than that of the $C_{3} $ correlation,
which appears to be as small as $2.09 \%$. 
This dependence of $C_{2} $ and $C_{4} $ on the direction 
constraints us to work with
directions close to the
preferred one, which strongly minimizes 
periodicity effects (see Sec. IV).

\begin{figure*}
\includegraphics[width=16cm]{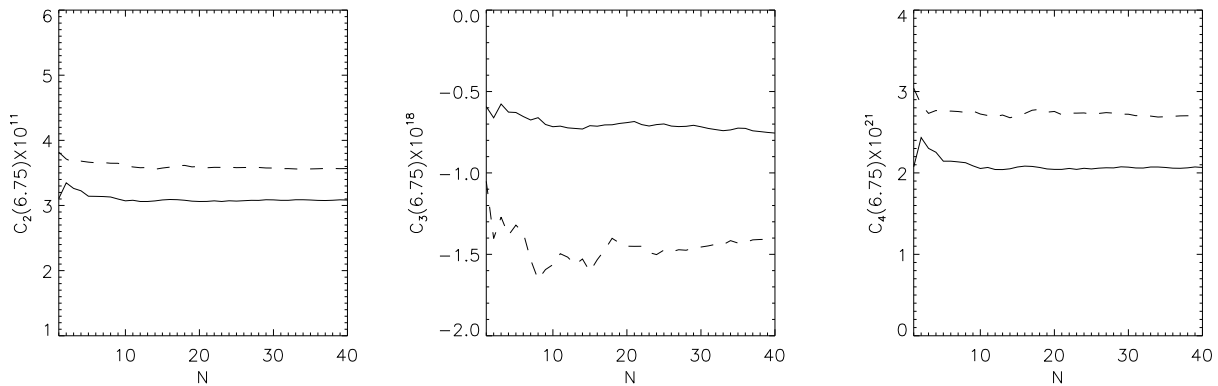}
\caption{\label{fig7}Same as Fig. \ref{fig6} for box sizes of 
$128 \ Mpc $ (dashed lines) and $256 \ Mpc $ (solid lines).}  
\end{figure*}

Finally, results of case 2 are compared with those of case 5 
to analyze the importance of the box size. Case
5 is identical to case 2 except for this size, which takes 
on the values 
$256 \ Mpc $  and $128 \ Mpc $ in cases 5 and 2, respectively.
Possible differences between 
cases 2 and 5 can be explained in two different ways: (i)
boxes of
$256 \ Mpc $ contain linear inhomogeneities more
extended than those involved in boxes of $128 \ Mpc $ and,
according to our discussion of Sec. III, linear inhomogeneities
with different sizes produce distinct lensing and,    
(ii) there is a residual effect of
periodicity, which would depend on box size for the fixed 
preferred direction used in our calculations (initially associated
to $128 \ Mpc $ boxes, see Sec. IV).
Results of case 5 are shown in the fifth
row of table~\ref{tbl-1}, and also in the solid lines of 
Fig. \ref{fig7} (where dashed lines correspond to
case 2).
Comparing the solid and dashed lines of Fig. \ref{fig7}, 
and taking into account that the comparation percentages of the pair
2-5 (forth row of table~\ref{tbl-2})
are $13.73 \%$ for $C_{2}$, $46 \%$ for $C_{3}$, and
$23 \%$ for $C_{4}$, we conclude that 
box size is important. Is that due to residual periodicity? 
The answer to this question is negative because the 
comparation percentage in $C_{3}$ due to periodicity has been 
proved to be very small ($2.09 \%$), whereas the corresponding 
percentage has appeared to be $46 \%$ when the box size has been
varied. Hence, 
the box size is important as a result of the presence 
of large scale inhomogeneities and, consequently, 
our best estimate of the correlations induced by nonlinear lensing 
corresponds to the smallest box size ($128 \ Mpc $) and the best resolution
($0.5 \ Mpc $) 
we have used (case 1). The comparation percentages of the pair 2-5 
are not to be confused with the error percentages in the correlations
of case 2, which would be expected to be smaller, perhaps about
a few per cent. The same has occurred for resolution, where 
the comparation percentages of the pair 2-3 (second row of table~\ref{tbl-2})
are much larger than the error percentages of case 2 (comparable to
the quantities in the first row of table~\ref{tbl-2}).

\section{Discussion}

A ray-tracing method to calculate correlations 
in lens deformations of CMB maps
has been designed, tested, and applied. 
It is based on the use of PM N-body simulations to evolve 
nonlinear cosmological structures. Although 
correlations are calculated
from the deformations associated to a reduced number of 
directions (no complete maps of deformations are created), 
numerical calculations are time consuming 
because a large enough number of  N-body simulations
must be performed to calculate good statistical averages (correlations).
The method is similar to that proposed by
Aliaga et al. \cite{ali02}
to study the Rees-Sciama effect, and it can be 
easily implemented starting from an N-body code. 

Firts of all, the main structures producing lens deformations in the
CMB sky have been identified. 
A qualitative 
estimation of the lens deformations produced by a representative 
set of linear and nonlinear cosmological structures has been used 
to achieve this identification. Various reasons have motivated  
the choice of the angular scale $\alpha = 6.75^{\prime}$
to compute correlations (see Sec. III). For this scale, it
has been proved that 
lens deformations are mainly produced by 
big and standard clusters, although there is 
a moderate contribution from 
linear inhomogeneities with diameters greater than $100 \ Mpc $.
This information has been crucial to choose the N-body simulations 
in such a way that
the effect of the linear inhomogeneities contained
in the simulation box is much smaller than that of the nonlinear 
structures. Using these simulations, the 
deviation field $\vec {\delta }$ produced by
nonlinear structures can be numerically estimated and, independently, 
the field $\vec {\delta }$
due to large scale linear inhomogeneities can be 
analytically calculated. 
Qualitative arguments suggest that 
box sizes between
$128 \ Mpc$ and $256 \ Mpc$ could be appropriate to achieve
the above requirements (Sec. III) and, then, 
experiments with simulations (see Sec. V) have been used to reach 
the following conclusions: (i) 
a size of $128 \ Mpc$ is preferable against larger sizes 
including extended linear structures (which would produce a
significant lensing) and, (ii)
resolutions of $0.5 \ Mpc$ and $1. \ Mpc $
lead to similar correlations; hence, results converge 
as resolution increases.
Is this convergence real or apparent? 
As it has been pointed out in  Sec. V,
the resolution of our simulations must be increased
to answer this question. How much should we 
increase the resolution? We should consider --at least-- 
a resolution of $0.25 \ Mpc$ and, perhaps another of 
$0.125 \ Mpc $. These resolutions lead to severe 
computational problems because many simulations are 
necessary for statistical purposes. 
Higher-resolution codes
--for example tree or  $P^{3}M$ codes-- could be 
more appropriate than our PM one. Fortunately, our simulations lead
to boxes containing nonlinear structures which allow us
to apply and test our ray-tracing procedure and, 
furthermore, our best resolution ($0.5 \ Mpc $) should lead to
a rather acceptable description of clusters,
because the central region, where the 
density is more peaked (distances $\sim 1 \ Mpc$ from the center),
is covered by about $2^{3} $ pairs of cells. We can state 
that our 
main goal (description of a new ray-tracing
procedure and estimations of nongaussianity from nonlinear
structures) has been reached, whereas results from higher
resolutions are out of the scope of this work, and
they will be developed in due time and presented elsewhere.
 
In our model, 
photons move in a periodic universe which is the repetition of 
the simulation box. We have verified that periodicity is important,
but photons do not feel periodicity when they move
along directions close to a preferred one (see Sec. IV); 
this fact constraints us
to calculate deformations for this type of directions. Fortunately,
these directions suffice 
to find the required correlations. It is worthwhile to emphasize some
novel aspects of our calculations: 
the basic equations are (\ref{funda1}) and (\ref{funda2});
the r.h.s of
Eq. (\ref{funda1}) has the form of a Fourier transform, and it
gives --at the same time-- the deviation field 
$\vec{\delta }(\vec{x}_{_{P}},\vec{n})$ for all the observers 
located in the nodes ( $\vec{x}_{_{P}}$ )
of the simulation grid. This fact has been probed to be relevant when 
averages are performed. The r.h.s. of Eq. 
(\ref{funda2}) is a time integral which 
can be analytically 
calculated after using the interpolation formula (\ref{interpol}), 
this procedure 
simplifies the calculations. 
The use of directions close to 
the preferred one is the 
last novel aspect deserving attention. 
For these directions and fixed boxes, 
photons enter the n-th box across a certain region  and
the (n+1)-th box through a
different independent zone; hence,
roto-traslations are not necessary to avoid
periodicity effects; furthermore, it is worthwhile to
emphasize that --without roto-traslations--
function $ \vec {\nabla}_{\bot } \phi$ is continuous
at the points where photons cross boundary boxes 
(periodicity). However,
if the boxes are moved (rototralations), 
discontinuities at these points are unavoidable and,
consequently, 
the function to be integrated
in Eq. (5) has artificial 
finite discontinuities which influence deviation calculations.
Perhaps these discontinuities do not produce any relevant total
effect when many boxes are crossed, but
potential problems with discontinuities are clearly surmounted 
by our method in the absence of roto-traslations, which
seem not to be either necessary or appropriate 
for us.

Since periodicity effects are prevented along any direction 
forming an angle of a few degrees with the preferred one, our 
method can be used: (1) to simulate 
squared maps with sizes of a few
degrees, for example, $4^{\circ} \times 4^{\circ}$ maps with
the preferred direction pointing towards their centers, and (2) to
compute angular correlations $C_{n}$ for 
$n>4 $. In previous sections, 
correlations $C_{3}$ and $C_{4}$ have been computed 
using sets of three and four directions close to the preferred  
one and, evidently, using the same methods and
other sets of these directions, the correlation $C_{6} $ 
and higher ones can be computed. 
On account of these considerations,
future applications of the proposed method are certainly promising.

Hereafter, we only make reference to 
the lens effect (or lens deformations) produced by nonlinear structures.
As it was predicted by Bernardeau \cite{ber97},
the lens effect does not create any 
significant $C_{3}$ correlation ($r_{3} \simeq 1.17 \times 10^{-4}$).
Correlations $C_{2}$ and $C_{4}$ appear to be 
weaker than those of the primary maps (low correlation level, see
Sec. V). The resulting ratio $3 C_{2}^{2}/C_{4} \simeq 1.43$ 
points out a deviation from Gaussianity. In the observational maps,
the lens deformations are superimposed to both the 
dominant primary anisotropies and other contributions 
described in the introduction (foregrounds and nonlinear effects).
Let us analyze the superposition of primary maps and 
lens deformations taking into account that:  (1) 
the correlation 
$C_{2}$  ($C_{4}$) 
of primary maps has appeared to be greater
than that of lens deformations by a
factor close to $1700$ ($4 \times 10^{6}$), (2) 
the correlation $C_{3}$ vanishes in primary maps
and it has been found to be negligible for deformations, and
(3) the two components under consideration 
are statistically independent. Simple calculations give 
the correlations of the superposition of components, which 
are almost identical to those of the primary maps.
Deviations with respect to Gaussianity 
are almost negligible in this superposition.
In the observational maps, there are other deviations due to 
the presence of other components, which would contribute to hide 
the small deviations associated to lens deformations. 

In future, our calculations could be repeated using other
N-body codes and resolutions,
the lens
effect due to linear structures contained in boxes with different
sizes could be studied in more detail; 
the present study could be extended to
other components of the observational CMB maps, in particular
to the Rees Sciama component \cite{ali02};
maps of lens deformations and other components could be simulated
--with our method-- and these maps (which would
have the true statistical properties at any order
of correlation) could be analyzed using various 
estimators of nonGaussianity proposed in the
literature \cite{tak01,ber98}; 
our simulations could be also used to study the
correlation between galaxy shear and CMB temperature distortions
\cite{van00} and, finally, using other N-body codes and
greater resolutions,
it would be worthwhile the study of very small (large)
angular scales 
($\ell $ values), for which, the $C_{\ell} $ coefficients
of the primary anisotropy are very small.
Some of these studies and other possible ones 
could be useful to improve on the                              
method proposed in this paper; nevertheless, 
a good version has been already
implemented and numerical codes are operative.

\begin{acknowledgments}
This work has been partially 
supported by the Spanish MCyT 
(project AYA2000-2045). VQ
is Ram\'on y Cajal Fellow of the Spanish Ministry of Science 
and Technology,
and PC thanks the same Ministry for a fellowship.
Calculations
were carried out on a SGI Origin 2000s at the Centro de Inform\'atica
de la Universitat de Val\`encia.
\end{acknowledgments}

\end{document}